\documentclass[prb,twocolumn,amssymb,showpacs,amsmath,nobibnotes,superscriptaddress,floatfix,aps]{revtex4}
\usepackage{graphicx}
\usepackage{bm}

\begin{document}
\title{Coexistence of the antiferromagnetic and superconducting order and its effect on spin dynamics in electron-doped high-$T_{c}$ cuprates }

\author{Cui-Ping Chen}
\affiliation{National Laboratory of Solid State of Microstructure
and Department of Physics, Nanjing University, Nanjing 210093,
China}
\author{Hong-Min Jiang}
\affiliation{National Laboratory of Solid State of Microstructure
and Department of Physics, Nanjing University, Nanjing 210093,
China}
\author{Jian-Xin Li}
\affiliation{National Laboratory of Solid State of Microstructure
and Department of Physics, Nanjing University, Nanjing 210093,
China}

\date{\today}

\begin{abstract}
In the framework of the slave-boson approach to the $t-t'-t''-J$
model, it is found that for electron-doped high-$T_c$ cuprates,
the staggered antiferromagnetic (AF) order coexists with
superconducting (SC) order in a wide doping level ranged from
underdoped to nearly optimal doping at the mean-field level. In
the coexisting phase, it is revealed that the spin response is
commensurate in a substantial frequency range below a crossover
frequency $\omega_{c}$ for all dopings considered, and it switches
to the incommensurate structure when the frequency is higher than
$\omega_{c}$. This result is in agreement with the experimental
measurements. Comparison of the spin response between the
coexisting phase and the pure SC phase with a
$d_{x^{2}-y^{2}}$-wave pairing plus a higher harmonics term
(DP+HH) suggests that the inclusion of the two-band effect is
important to consistently account for both the dispersion of the
spin response and the non-monotonic gap behavior in the
electron-doped cuprates.

\end{abstract}


\maketitle

\section{Introduction}
The pairing symmetry of the hole-doped high-$T_{c}$ superconductors
is generally believed to have the dominant $d_{x^{2}-y^{2}}$-wave
pairing. However, the pairing symmetry of the electron-doped
high-$T_{c}$ superconductors is still under
debate.~\cite{1,2,3,4,5,6,7,8} While no consensus has been reached
yet, more and more recent experimental results have suggested that
the order parameter of electron-doped cuprates is likely to have a
dominant $d_{x^{2}-y^{2}}$-wave pairing symmetry,~\cite{3,4,5,6,7,8}
and with an unusual non-monotonic gap function.

Although various explanations have been proposed to account for the
non-monotonic behavior, they can generally be categorized into two
scenarios.~\cite{5,7,9,10,11,12,13,14,15,16,17,18} One is to extend
the superconducting (SC) gap out of the simple
$d_{x^{2}-y^{2}}$-wave via the inclusion of higher harmonics
term.~\cite{5,7,9,10,11,12,13} From theoretical perspective, the
non-monotonic $d_{x^{2}-y^{2}}$-wave gap appears under the
assumption that the $d_{x^{2}-y^{2}}$-wave pairing is caused by the
interaction with the continuum of overdamped antiferromagnetic (AF)
spin fluctuations. In this scenario, the non-monotonic gap behavior
is described by the combination effect of a $d_{x^{2}-y^{2}}$-wave
paring plus a higher harmonics term (DP+HH). Therefore, it is an
intrinsic property of the SC state regardless of the presence of the
AF order, and a simple one-band model is used to reproduce the
non-monotonic gap behavior. The other argues that the non-monotonic
behavior is the outcome of the coexistence of the AF and the SC
orders.~\cite{14,15,16,17,18} This scenario assumes that the AF
order disguises the $d_{x^{2}-y^{2}}$-wave character of SC gap. When
the AF order is introduced, the resulting quasiparticle (QP)
excitation can be gapped by both orders and behaves to be
non-monotonic, although the SC gap itself is monotonic. The scenario
gained support from the angle-resolved photoemission spectra (ARPES)
measurements, where two inequivalent Fermi pockets around $(\pi, 0)$
and $(\pi/2, \pi/2)$ have been detected.~\cite{19,20} This phenomena
is well explained in terms of the $\mathbf{k}$-dependent
band-folding effect associated with an AF order which splits the
band into upper and lower branches,~\cite{14,20,21,22} leading to
the two-band and/or two-gap model.

Recently, neutron scattering experiments in electron-doped cuprates
have revealed that the spin response is commensurate in a
substantial frequency range below a crossover frequency $\omega_{c}$
,\cite{23,24,25,26,27,28} which constitutes a distinct difference
from the widely studied hour-glass dispersion in the hole-doped
cuprates.\cite{29} Although, both scenarios mentioned above can
account for the non-monotonic gap behavior of the electron-doped
cuprates, a comparative study on the spin dynamics between the two
scenarios is deserved to demonstrate the possible differences and
therefore serve to select the reasonable model for electron-doped
high-$T_{c}$ cuprates.

In this paper, we investigate the spin dynamics in the coexisting
phase of the AF  and the $d_{x^{2}-y^{2}}$-wave SC orders. The
calculation is based on a self-consistent determination of the QP
dispersion, the AF order and the SC gap at the slave-boson
mean-field level of the $t-t'-t''-J$ model. It is shown that the AF
and SC orders compete and coexist in a substantial doping range in
the underdoped regime. The spin response is commensurate below a
crossover frequency $\omega_{c}$ for all dopings considered, and it
becomes incommensurate when the frequency is higher than
$\omega_{c}$. This result is qualitatively consistent with
experiments.~\cite{23,24,25,26,27,28} While in the framework of the
pure SC state with $d_{x^{2}-y^{2}}$-wave and/or DP+HH,
~\cite{30,31,32}, though an extended region of a commensurate spin
fluctuation also exists, it evolves into an incommensurate spin
fluctuation at low frequencies, which is not consistent experiments.
Therefore, our result suggests that the inclusion of the two-band
effect resulting from the coexisting AF and SC orders is important
to consistently account for both the spin dynamics and the
non-monotonic gap behavior in the electron-doped cuprates.

The paper is organized as follows. In Sec. II, we introduce the
theoretical model and carry out the analytical calculations. In Sec.
III, we present the numerical results with some discussions.
 Finally, we present the conclusion in Sec. IV.

\section{THEORETICAL MODEL}
The Hamiltonian of the two dimensional $t-t'-t''-J$ model on a
square lattice is written in the form,
\begin{widetext}
\begin{eqnarray}
H &=&
-t\sum_{<ij>,\sigma}(c^{\dagger}_{i\sigma}c_{j\sigma}+H.c.)-t_{1}\sum_{<ij>_{2},\sigma}(c^{\dagger}_{i\sigma}c_{j\sigma}+H.c.)-t_{2}\sum_{<ij>_{3},\sigma}(c^{\dagger}_{i\sigma}c_{j\sigma}+H.c.)\nonumber\\&&+J\sum_{<ij>}(\mathbf{S}_i\cdot\mathbf{S}_j-\frac{1}{4}n_{i}n_{j})-\mu_{0}\sum_{<i>,\sigma}c^{\dagger}_{i\sigma}c_{i\sigma}.
\end{eqnarray}
\end{widetext}
Where the summations  $<ij>$, $<ij>_{2}$, $<ij>_{3}$ run over the
nearest-neighbor(n$\cdot$n), the next-n$\cdot$n, and the
third-n$\cdot$n pairs respectively, $\mathbf{S}_i$ is the spin on
site $i$. This Hamiltonian can be used to model both hole-doped and
electron-doped systems after a particle-hole transformation. For
electron doping, one has $t$$<$$0$, $t_{1}$$>$$0$ and $t_{2}$$<$$0$.
The slave-boson mean-field theory (SBMFT) is used to decouple the
electron operators $c_{i\sigma}$ to  bosons $b_i$ carrying the
charge and fermions $f_{i\sigma}$ representing the spin. Then, the
local constraint
$b^{\dagger}_ib_i+\sum_{i\sigma}f^{\dagger}_{i\sigma}f_{i\sigma}$=$1$
is satisfied averagely at the mean-field (MF) level. We choose the
spinon pairing order
$\Delta_{ij}$=$<f_{i\uparrow}f_{j\downarrow}-f_{i\downarrow}f_{j\uparrow}>$=$\pm\Delta$
, where $\Delta_{ij}=\Delta(-\Delta)$ for bond $<ij>$ along $x(y)$
direction, the uniform bond order
$\chi_{ij}$=$\sum_{\sigma}<f^{\dagger}_{i\sigma}f_{j\sigma}>$=$\chi$,
the AF order
$<f_{i\uparrow}^{\dag}f_{i\uparrow}-f_{i\downarrow}^{\dag}f_{i\downarrow}>/2=(-1)^{i}m$,
and replace $b_i$ by $<b_i>$=$\sqrt{x}$ due to boson condensation.
After the Fourier transformation, the mean-field (MF) Hamiltonian
can be written in the Nambu representation,
\begin{equation}
H=\sum_{\mathbf{k}}C^{\dag}(\mathbf{k})\hat{A}(\mathbf{k})C(\mathbf{k})+2NJ(\chi^{2}+m^{2}+\Delta^{2}/2)-N\mu,
\end{equation}
here, the Nambu operator
$C^{\dag}(\mathbf{k})=(f_{\mathbf{k}\uparrow}^{\dag},f_{\mathbf{k+Q}\uparrow}^{\dag},
f_{\mathbf{-k}\downarrow},f_{\mathbf{-k-Q}\downarrow})$, and
\begin{eqnarray}
\hat{A}(\mathbf{k})= \left(
\begin{array}{c c c c}
{\epsilon_{\mathbf{k}}}&{-2Jm}&{-J\Delta_{\mathbf{k}}}&0\\{-2Jm}&{\epsilon_{\mathbf{k+Q}}}&0&{J\Delta_{\mathbf{k}}}\\{-J\Delta_{\mathbf{k}}}&0&{-\epsilon_{\mathbf{k}}}&{-2Jm}\\
0&{J\Delta_{\mathbf{k}}}&{-2Jm}&{-\epsilon_{\mathbf{k+Q}}}
\end{array}\right),
\end{eqnarray}
where, $\epsilon_{\mathbf{k}}=(-2tx-J\chi)(\cos k_{x}+\cos
k_{y})-4t_{1}x\cos k_{x}\cos
k_{y}-2t_{2}x(\cos2k_{x}+\cos2k_{y})-\mu$ and
$\Delta_{\mathbf{k}}=\Delta(\cos k_{x}-\cos k_{y})$. $\mu$ is the
renormalized chemical potential, $N$ is the total number of lattice
sites, and $\mathbf{Q}=(\pi,\pi)$ is the AF momentum. Note that the
wave vector $\mathbf{k}$ is restricted to the magnetic Brillouin
zone (MBZ) in all follows.

Diagonalizing of the Hamiltonian (2) by an unitary matrix
$\hat{U}(\mathbf{k})$ leads to four energy bands
$E_{1}(\mathbf{k})=E^{+}_{\mathbf{k}}$,
$E_{2}(\mathbf{k})=E^{-}_{\mathbf{k}}$,
$E_{3}(\mathbf{k})=-E^{-}_{\mathbf{k}}$,
$E_{4}(\mathbf{k})=-E^{+}_{\mathbf{k}}$, with
\begin{equation}
E^{\pm}_{
\mathbf{k}}=\sqrt{(\xi_{\mathbf{k}}^{\pm})^{2}+(J\Delta_{\mathbf{k}})^{2}},
\end{equation}
where
$\xi_{\mathbf{k}}^{\pm}=\epsilon^{+}_{\mathbf{k}}\pm\sqrt{(\epsilon^{-}_{\mathbf{k}})^{2}
+4J^{2}m^{2}}$ with
$\epsilon^{\pm}_{\mathbf{k}}=(\epsilon_{\mathbf{k}}\pm\epsilon_{\mathbf{k+Q}})/2$.
And the free energy is written down (Boltzmann constant $k_{B}=1$),
\begin{equation}
F=-2T\sum_{\mathbf{k},\nu=\pm}\ln[2\cosh(\frac{E_{\mathbf{k}}^{\nu}}{2T})]-\mu
N+2NJ(\chi^{2}+m^{2}+\Delta^{2}/2).
\end{equation}
The MF order parameters $\chi$, $\Delta$, $m$ and the chemical
potential $\mu$ for different dopings $x$ can be calculated from the
self-consistent equations obtained by $\partial F/\partial \chi=0$,
$\partial F/\partial \Delta=0$, $\partial F/\partial m=0$, and
$\partial F/\partial \mu=-N(1-x)$, respectively. The magnitudes of
the parameters are chosen as $t$=$-3.0J$, $t_{1}$=$1.02J$,
$t_{2}$=$-0.51J$ and $J$=$100$ meV to model the Fermi surface
observed in ARPES experiment.\cite{19,20}

Then, the bare spin susceptibility (transverse) is given by,
\begin{equation}
\chi^{\pm}_{0}(\mathbf{q},\mathbf{q}^{'},\tau)=\frac{1}{N}<S^{+}_{\mathbf{q}}
(\tau)S^{-}_{-\mathbf{q}'}(0)>_{(0)},
\end{equation}
where $<\cdots>_{(0)}$ means thermal average over the eigenstates of
$H$,
$S^{+}_{\mathbf{q}}=\sum_{k}f^{+}_{\mathbf{k+q}\uparrow}f_{\mathbf{k}\downarrow}$
is the spin operator. Considering that $\mathbf{k}$ is restricted to
the MBZ, an explicit calculation shows that the spin susceptibility
should be expressed in the following matrix form,
\begin{eqnarray}
\ \hat{\chi}_{0}^{\pm}(\mathbf{q},\omega)= \left(
\begin{array}{c c}
{\chi_{0}^{\pm}(\mathbf{q},\mathbf{q},\omega)}&{\chi_{0}^{\pm}(\mathbf{q},\mathbf{q+Q},\omega)}\\{\chi_{0}^{\pm}(\mathbf{q+Q},\mathbf{q},\omega)}&{\chi_{0}^{\pm}(\mathbf{q+Q},\mathbf{q+Q},\omega)}
\end{array}\right),
\end{eqnarray}
where, the nondiagonal correlation function $\chi_{0}^{\pm}$ with
$\mathbf{q}'=\mathbf{q+Q}$ arises due to the umklapp process. The
matrix elements of the bare spin susceptibility, which come from the
particle-hole $(p-h)$ excitations, are given by,
\begin{widetext}
\begin{eqnarray}
\chi_{0}^{\pm}(\mathbf{q},\omega)_{\eta\eta'}& = &
\frac{1}{N}\sum_{i,j=1}^{2}\sum_{m,n=1}^{2}\sum_{\mathbf{k}}
[a_{1}\frac{f(E_{m}(\mathbf{k}))-f(E_{n}(\mathbf{k+q}))}{\omega+E_{n}(\mathbf{k+q})-E_{m}(\mathbf{k})+i\Gamma}
+a_{2}\frac{f(E_{n}(\mathbf{k+q}))-f(E_{m}(\mathbf{k}))}{\omega-E_{n}(\mathbf{k+q})+E_{m}(\mathbf{k})+i\Gamma}
\nonumber\\&&+b_{1}\frac{1-f(E_{m}(\mathbf{k}))-f(E_{n}(\mathbf{k+q}))}
{\omega+E_{n}(\mathbf{k+q})+E_{m}(\mathbf{k})+i\Gamma}
+b_{2}\frac{f(E_{m}(\mathbf{k}))+f(E_{n}(\mathbf{k+q}))-1}{\omega-E_{n}(\mathbf{k+q})-E_{m}(\mathbf{k})+i\Gamma}],
\end{eqnarray}
\end{widetext}
where, $f(E_\mathbf{\mathbf{k}})$ is the Fermi
function and
\begin{widetext}
\begin{eqnarray}
a_{1}&=&
U_{in}^{*}(\mathbf{k+q})U_{(j+\eta'-\eta)n}(\mathbf{k+q})U_{im}(\mathbf{k})U_{jm}^{*}(\mathbf{k})
+U_{in}^{*}(\mathbf{k+q})U_{(j+2)n}(\mathbf{k+q})U_{im}(\mathbf{k})U_{(j+2+\eta'-\eta)m}^{*}(\mathbf{k}),
\nonumber\\
a_{2}&=&
U_{(i+2)n}(\mathbf{k+q})U_{(j+2+\eta'-\eta)n}^{*}(\mathbf{k+q})U_{(i+2)m}^{*}(k)U_{(j+2)m}(\mathbf{k})
+U_{(i+2)n}(\mathbf{k+q})U_{jn}^{*}(\mathbf{k+q})U_{(i+2)m}^{*}(\mathbf{k})U_{(j+\eta'-\eta)m}(\mathbf{k}),
\nonumber\\
b_{1}&=&
U_{in}^{*}(\mathbf{k+q})U_{(j+\eta'-\eta)n}(\mathbf{k+q})U_{(i+2)m}^{*}(\mathbf{k})U_{(j+2)m}(\mathbf{k})
-U_{in}^{*}(\mathbf{k+q})U_{(j+2)n}(\mathbf{k+q})U_{(i+2)m}^{*}(\mathbf{k})U_{(j+\eta'-\eta)m}(\mathbf{k}),
\nonumber\\
b_{2}&=&
U_{(i+2)n}(\mathbf{k+q})U_{(j+2+\eta'-\eta)n}^{*}(\mathbf{k+q})U_{im}(\mathbf{k})U_{jm}^{*}(\mathbf{k})
-U_{(i+2)n}(\mathbf{k+q})U_{jn}^{*}(\mathbf{k+q})U_{im}(\mathbf{k})U_{(j+2+\eta'-\eta)m}^{*}(\mathbf{k}).
\end{eqnarray}
\end{widetext}

The renormalized spin susceptibility due to the spin fluctuations is
obtained via the random-phase approximation (RPA),
\begin{equation}
\hat{\chi}^{\pm}(\mathbf{q},\omega)=\frac{\hat{\chi}_0^{\pm}(\mathbf{q},\omega)}{\hat{1}+\alpha
\hat{J}_{q}\hat{\chi}_0^{\pm}(\mathbf{q},\omega)},
\end{equation}
where,
\begin{eqnarray}
 \hat{J}_{q}= \left(
\begin{array}{c c}
{J(\mathbf{q})}&{0}\\{0}&{J(\mathbf{q+Q})}
\end{array}\right)
\end{eqnarray}
with $J(\mathbf{q})=J(\cos q_{x}+\cos q_{y})$. In the coexisting
phase of the AF order and SC order, $\alpha$ is taken as 1. As for
the pure SC state with DP+HH, we choose a slightly small
$\alpha=0.72$, the criteria for choosing $\alpha$ is to set the AF
instability at $x=0.12$. The parameter $\Gamma$=$0.04J$ is
introduced to account for the QP damping rate which comes from the
scattering off other fluctuations that are not included here.

\section{NUMERICAL RESULTS AND DISCUSSION}
In Fig. 1, we show the MF parameters $\chi$, $m$ and $\Delta$  as a
function of doping $x$. For a comparison, we also show the doping
dependence of the MF SC gap $\Delta_{1}$ obtained without
considering the AF order by setting $m=0$. It is seen that the
staggered magnetization $m$ decreases with increasing doping $x$,
and goes sharply to zero at $x\approx0.16$, which implies a phase
transition from the antiferromagnetism (AFM) phase to the
paramagnetic phase. The SC order parameter, on the other hand,
increases its value initially up to an optimal doping level, and
then decreases upon further doping, forming a generic SC
dome.~\cite{33} However, the SC order parameter $\Delta_{1}$ without
the inclusion of the AF order exhibits a monotonic decrease with
doping, which deviates obviously from the experimental observations.
Furthermore, the SC order parameter $\Delta$ with an AF order shows
a noticeable suppression compared to $\Delta_{1}$, exhibits a
competitive character with the AF order. But, they also coexist in a
substantial doping range. The MF phase diagram also shows that the
optimal doping is rather low compared to that deduced from the
experiments. This may be due to the fact the SBMFT includes only the
MF value of the order parameters and treats the no-double occupancy
on the average. However, the similarity of the phase diagram
obtained by the SBMFT to that of the variational quantum-cluster
theory~\cite{16,22} validates the SBMFT as a low energy effective
theory. Here our aim is to use the MF theory as an effective model
to study the effect of the AF order on the spin dynamics, and then
to compare the two-band and/or two-gap model with the simple
one-band model. Therefore, the relatively simple SBMFT is
qualitatively enough for our purpose. We note that a similar phase
diagram has been obtained before.~\cite{15}

The doping dependence of the renormalized spin susceptibility
Im$\chi(\mathbf{q},\omega)$ at a low frequency $\omega=0.04J$ in the
coexisting phase is presented in Fig. 2. In this figure, it is found
that the low-energy excitations exhibit commensurate peaks for all
$x$, which consists with the experiments well~\cite{27}. The inset
shows the spin susceptibility Im$\chi(\mathbf{q},\omega)$ at doping
$x=0.15$ in the pure SC state with DP+HH which is used to reproduce
a non-monotonic SC gap behavior. One can see that the spin response
is incommensurate at low frequency without considering the AF order.

Detailed frequency dependence of the spin response in the coexisting
phase and the pure SC phase with DP+HH at doping $x=0.15$ are shown
in Figs. 3(a) and 3(b), and Figs. 3(c) and 3(d), respectively. The
difference in the low frequency regime of the two phases is more
evident here. The spin fluctuation is commensurate in a substantial
frequency range below a crossover frequency $\omega_{c}\approx0.52J$
and down to the lowest frequency considered in the coexisting phase,
and it switches to be incommensurate when the frequency is higher
than $\omega_{c}\approx0.52J$ [Figs. 3(a) and 3(b)]. This feature
agrees with the neutron-scattering measurements on electron-doped
cuprates that have been reported recently.~\cite{23} While for the
pure SC phase with DP+HH, the spin response is incommensurate at low
frequency, then it switches to be commensurate within the
intermediate frequency range, and becomes incommensurate again at
higher frequency.~\cite{31} These results can be summarized in the
intensity plot of the imaginary part of the renormalized spin
susceptibility Im$\chi(\mathbf{q},\omega)$ as a function of
frequency and momentum along $(\pi,q_{y})$ direction, as shown in
Fig. 4. In the figure, the solid line indicating the peak position
is the dispersion of spin excitations. The commensurate spin
fluctuation prevails below $\omega_{c}$ for the coexisting system
[Fig. 4(a)]. For the pure SC phase with DP+HH, the dispersion shows
a hourglass-like behavior [Fig. 4(c)], which is similar to the
hole-doped one, and does not consistent with the experiments on
electron-doped cuprates.~\cite{23}

In the presence of the AF order, the energy band of quasiparticles
is split into two bands. Therefore, the particle-hole excitations
that contributed to the spin susceptibility are composed of two
kinds of excitations, the intra-band and the inter-band excitations.
In Fig. 5, we present the results for the bare spin susceptibility
$\chi_{0}(\mathbf{q},\omega)$ (without the RPA correction) coming
from the intra-band and the inter-band contributions, respectively.
Figs. 5(a1) and 5(a2) denote the imaginary part of
$\chi_{0}(\mathbf{q},\omega)$, Figs. 5(b1) and 5(b2) the real part.
One obvious feature is that, the intra-band contribution is zero at
the AF momentum $\mathbf{Q}=(\pi,\pi)$, leading to the
incommensurate spin response. It results from the fact that the
coherence factor in the spin susceptibility due to the intra-band
excitations, $1-[(2Jm)^{2}-\varepsilon_{\mathbf{k+q}}
\varepsilon_{\mathbf{k}}]/[{\sqrt{\varepsilon_{\mathbf{k+q}}^{2}+(2Jm)^{2}}
\sqrt{\varepsilon_{\mathbf{k}}^{2}+(2Jm)^{2}}}]$ [where,
$\varepsilon_{\mathbf{k}}=(-2tx-J\chi)(\cos k_{x}+\cos k_{y})$] is
zero at $\mathbf{Q}$. While, the inter-band contribution is
commensurate for all frequencies. At low frequencies, the inter-band
excitations have a larger contribution to the spin susceptibility
than the intra-band excitations, so the spin fluctuation is
commensurate. However, with the increase of frequency, the intensity
of Im$\chi_{0}(\mathbf{q},\omega)$ due to the intra-band
contributions increases more rapidly than the inter-band
contribution. As a result, the spin fluctuation switches from a
commensurate to an incommensurate structure.

\section{conclusion}
In this paper, we have investigated the spin dynamics in the
electron-doped cuprates in the coexisting phase of the
$d_{x^{2}-y^{2}}$-wave SC and AF orders, and compared the results
with that in the dominant $d_{x^{2}-y^{2}}$-wave phase with a higher
harmonics term. In the coexisting phase, we found that the spin
response is commensurate in a substantial frequency range below a
crossover frequency $\omega_{c}$ for all dopings considered, and it
switches to be incommensurate when the frequency is higher than
$\omega_{c}$. The theoretical calculations are shown to be in good
agreement with the experimental measurements. However, in the
dominant $d_{x^{2}-y^{2}}$-wave phase with a higher harmonics term,
the dispersion is just like that of the hole-doped one, namely
exhibits a hourglass-like dispersion, which is not consistent with
experiments. Thus, our result suggests that the inclusion of the
two-band effect is important to consistently account for both the
dispersion of the spin response and the non-monotonic gap behavior
in the electron-doped cuprates.

\begin{acknowledgments}
This work was supported by the National Natural Science Foundation
of China (10525415), the Ministry of Science and Technology of China
(973 project Grants Nos.2006CB601002,2006CB921800), and the China
Postdoctoral Science Foundation (Grant No. 20080441039).
\end{acknowledgments}

\vspace*{-.2cm}

\begin{figure}[htb]
\begin{center}
\vspace{-.2cm}
\includegraphics[width=200pt,height=180pt]{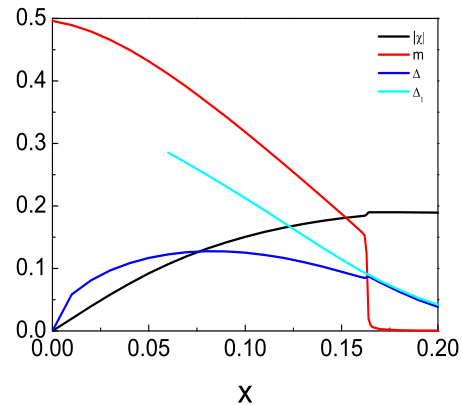}
\caption{(Color online) Mean-field phase diagram for $t-t'-t''-J$
model, where $\Delta_{1}$ is the superconducting order parameter
without considering the AF order. The model parameters are taken as:
$t=-3.0J$, $t'=1.02J$, $t''=-0.51J$.}\label{fig1}
\end{center}
\end{figure}

\vspace*{-.2cm}

\begin{figure}[htb]
\begin{center}
\vspace{-.2cm}
\includegraphics[width=260pt,height=230pt]{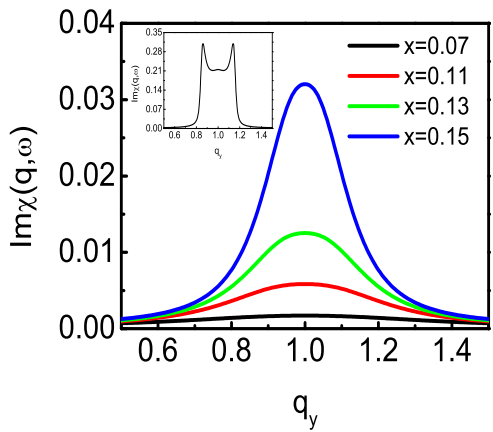}
\caption{(Color online) Doping dependence of
Im$\chi(\mathbf{q},\omega)$ in the coexisting phase of the AF and SC
order at low frequency $\omega=0.04J$. The momentum is scanned along
$(\pi ,q_{y})$. The inset shows Im$\chi(q,\omega)$ at $\omega=0.04J$
for the pure SC state with DP+HH at doping $x=0.15$. }\label{fig2}
\end{center}
\end{figure}

\vspace*{-.2cm}
\begin{figure}[htb]
\begin{center}
\vspace{-.2cm}
\includegraphics[width=280pt,height=270pt]{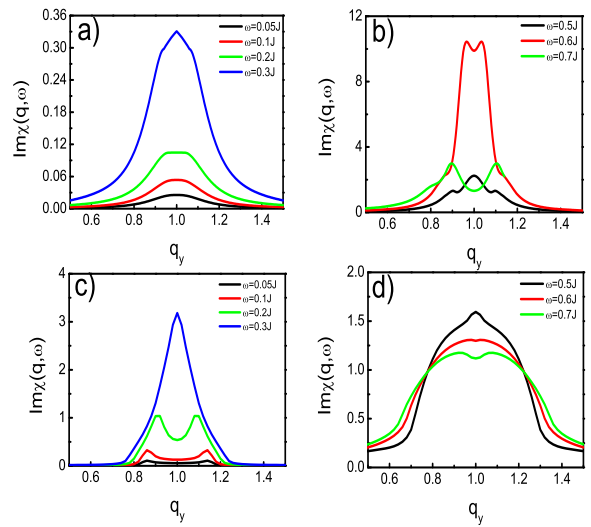}
\caption{(Color online) Frequency dependence of
Im$\chi(\mathbf{q},\omega)$ at doping $x$=$0.15$.  The momentum is
scanned along $(\pi,q_{y})$. (a) and (b) are in the coexistence of
AF and SC state. (c) and (d) are in the pure SC state with
DP+HH.}\label{fig4}
\end{center}
\end{figure}

\vspace*{-.2cm}
\begin{figure}[htb]
\begin{center}
\vspace{-.2cm}
\includegraphics[width=290pt,height=200pt]{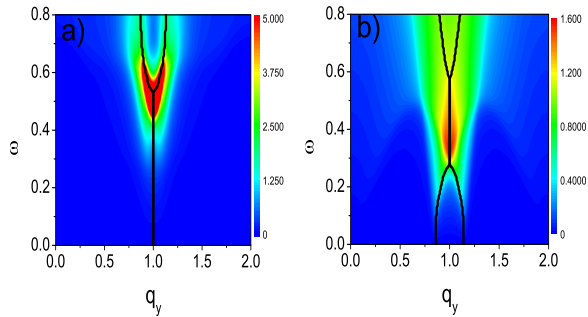}
\caption{(Color online) Intensity plot of
Im$\chi(\mathbf{q},\omega)$ as a function of frequency $\omega$ and
momentum $\mathbf{q}$ at doping $x$=$0.15$. The momentum is scanned
along $(\pi,q_{y})$. The solid line is the peak position. (a) is in
the coexistence of AF and SC state and (b) in the pure SC state with
DP+HH. }\label{fig3}
\end{center}
\end{figure}

\vspace*{-.2cm}
\begin{figure}[htb]
\begin{center}
\vspace{-.2cm}
\includegraphics[width=260pt,height=240pt]{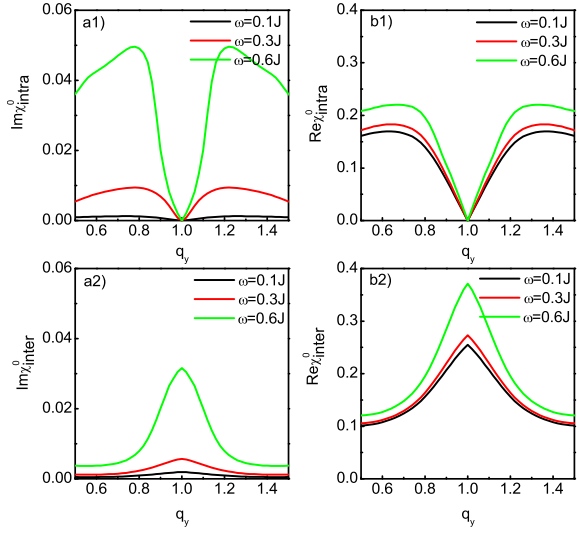}
\caption{(Color online) Frequency dependence of the intra- and
inter-band contributions to the bare spin susceptibility
$\chi_{0}(\mathbf{q},\omega)$ [(a1) and (a2) denote the imaginary
part, (b1) and (b2) the real part] in the coexistence of AF and SC
state at doping $x$=$0.15$. (a1) and (b1) show the intra-band
contributions , and (a2) and (b2) the inter-band
contribution.}\label{fig5}
\end{center}
\end{figure}

\end{document}